# The Integral Field Spectrograph for the Gemini Planet Imager

James E. Larkin[*a], Jeffrey K. Chilcote[a], Theodore Aliado[a], Brian J. Bauman[b], George Brims[a], John M. Canfield[a], Andrew Cardwell[c], Daren Dillon[d], René Doyon[e], Jennifer Dunn[f], Michael P. Fitzgerald[a], James R. Graham[g], Stephen Goodsell[c], Markus Hartung[c], Pascale Hibon[c], Patrick Ingraham[h,e], Christopher A Johnson[a], Evan Kress[a], Quinn M. Konopacky[i], Bruce A. Macintosh[b,h], Kenneth G. Magnone[a], Jerome Maire[i], Ian S. McLean[a], David Palmer[b], Marshall D. Perrin[j], Carlos Quiroz[c], Fredrik Rantakyrö[c], Naru Sadakuni[c], Leslie Saddlemyer[f], Andrew Serio[c], Simon Thibault[k], Sandrine J. Thomas[l,d], Philippe Vallee[e], Jason L. Weiss[a]

[a]Department of Physics & Astronomy, University of California, Los Angeles, CA, 90095, USA
[b]Lawrence Livermore National Laboratory, 7000 East Ave., Livermore, CA 94040, USA
[c]Gemini Observatory, Casilla 603, La Serena, Chile
[d]UARC, UC Santa Cruz, Santa Cruz CA 95064 USA
[e]Department de Physique, Université de Montréal, Montréal QC H3C 3J7, Canada
[f]National Research Council of Canada Herzberg, 5071 West Saanich Road, Victoria, BC V9E 2E7, Canada
[g]Department of Astronomy, UC Berkeley, Berkeley CA, 94720, USA
[h]Kavli Institute for Particle Astrophysics and Cosmology, Stanford University, Stanford, CA 94305, USA
[i]Dunlap Institute for Astrophysics, University of Toronto, 50 St. George St, Toronto, ON, M5S 3H4, Canada
[j]Space Telescope Science Institute, 3700 San Martin Drive, Baltimore, MD 21218, USA
[k]Université Laval, Departement de Physique, Québec, QC, G1V 0A6, Canada
[l]NASA Ames Research Center, Moffett Field, CA 94035, USA

## ABSTRACT

The Gemini Planet Imager (GPI) is a complex optical system designed to directly detect the self-emission of young planets within two arcseconds of their host stars. After suppressing the starlight with an advanced AO system and apodized coronagraph, the dominant residual contamination in the focal plane are speckles from the atmosphere and optical surfaces. Since speckles are diffractive in nature their positions in the field are strongly wavelength dependent, while an actual companion planet will remain at fixed separation. By comparing multiple images at different wavelengths taken simultaneously, we can freeze the speckle pattern and extract the planet light adding an order of magnitude of contrast. To achieve a bandpass of 20%, sufficient to perform speckle suppression, and to observe the entire two arcsecond field of view at diffraction limited sampling, we designed and built an integral field spectrograph with extremely low wavefront error and almost no chromatic aberration. The spectrograph is fully cryogenic and operates in the wavelength range 1 to 2.4 microns with five selectable filters. A prism is used to produce a spectral resolution of 45 in the primary detection band and maintain high throughput. Based on the OSIRIS spectrograph at Keck, we selected to use a lenslet-based spectrograph to achieve an rms wavefront error of approximately 25 nm. Over 36,000 spectra are taken simultaneously and reassembled into image cubes that have roughly 192x192 spatial elements and contain between 11 and 20 spectral channels. The primary dispersion prism can be replaced with a Wollaston prism for dual polarization measurements. The spectrograph also has a pupil-viewing mode for alignment and calibration.

## 1. INTRODUCTION

The spectrograph within the Gemini Planet Imager[1] instrument is a cryogenic integral field spectrograph (IFS) utilizing a monolithic lenslet array to partition the field of view (2.7 arcsec on a side) at the infrared diffraction limit of the Gemini Telescope (platescale is 14.14±0.01 milliarcsec per lenslet). Much of its design is based on the OSIRIS spectrograph[2] which was the first cryogenic integral field spectrograph based on a lenslet array and also designed to Nyquist sample the diffraction limit of a large telescope (Keck in the case of OSIRIS). The design is also similar to Project 1640[3] and there was strong collaboration between GPI and Project 1640. All of the lenslet based spectrographs rely on the ground breaking designs from Roland Bacon such as the optical Tiger Instrument[4]. The diffraction limited nature of the GPI, Project-1640 and OSIRIS specrtrographs leads to diffraction effects on the pupil images formed by the lenslets and requires significantly faster optics compared to seeing limited instruments. The extreme magnification of the Gemini Planet Imager and high level of contrast require very precise relative positioning of the optical elements with respect to outside reference surfaces and extremely low wavefront error. The wavefront error requirement (<25nm rms) in particular drove the selection of the lenslet-based design since only optical elements in front of the lenslets contribute to image degradation. A lenslet design is also straightforward to expand to the 36,000+ spectra needed to cover the field of view of the instrument. The wavelength range of the five spectral modes are shown in Table 1.

*larkin@astro.ucla.edu; phone 1-310-825-9790; fax 1-310-206-7254

Table 1 -Filter half power points, spectral resolutions including variations across the field and the number of pixels in a typical single spectrum along the dispersion axis.

| Filter Name | ½ power wavelengths | %bandpass | Spectral Resolution($\lambda/\Delta\lambda$) | # spectral pixels |
|---|---|---|---|---|
| Y | 0.95-1.14μm | 18% | 34-36 | 12-13 |
| J | 1.12-1.35 | 19% | 35-39 | 13-15 |
| H | 1.50-1.80 | 18% | 44-49 | 16-18 |
| K1 | 1.9-2.19 | 14% | 62-70 | 18-20 |
| K2 | 2.13-2.4 | 12% | 75-83 | 18-20 |

The Gemini Planet Imager(GPI)[1] is one of a new generation of dedicated high contrast adaptive optics systems with the goal of surveying young stellar systems and directly imaging self-luminous jovian planets. It can also study debris disks around young stars and has both spectroscopic and polarization modes. GPI is led by Bruce Macintosh (Lawrence Livermore National Laboratory and now Stanford University) and is an international collaboration of eight institutions. The instrument has recently been deployed at the Gemini South Observatory and has seven major sub-components: adaptive optics (AO) system[5], calibration interferometer (CAL)[6], coronagraph masks[7], cryogenic integral field spectrograph (IFS)[8], opto-mechanical superstructure (OMSS), top level software[9] and a data reduction pipeline[10][12]. The AO system is composed of a low spatial frequency, high stroke, 11x11 actuator deformable mirror, and a 64x64 Microelectromechanical System (MEMS) deformable mirror from Boston Micromachines (BMC). The AO system is innovative in that it includes a spatial filter to prevent aliasing and produce a square, dark region very close to the star[13]. This paper presents the integral field spectrograph (IFS). Overall responsibility for the IFS development was at UCLA (PI Larkin) where a related integral field spectrograph (OSIRIS) was constructed and delivered to Keck in early 2005 and where several other facility class instruments have been built since 1989. The spectrograph optics after the lenslet array were managed as a subcontract to the University Laval (ULaval/ImmerVision) and University of Montreal (UdeM) under the leads of Thibault and Doyon using heritage from previous near infrared systems including WIRCAM[14] for CfHT.

## 2. OPTICAL DESIGN

The IFS optical system can be broken into several functional pieces (see Figure 1 for schematic representation) that will be discuss in the sub-sections below. Initially, a collimated beam produced by the CAL unit enters the spectrograph through an infrared transmissive window that serves as a vacuum seal. A wheel of 9 cold Lyot stops and one blank are located in the pupil plane. For alignment and calibration a mirror can be inserted after the pupil plane and a lens can reimage the pupil plane onto a small commercial infrared camera. For science operations, the fold mirror is removed and a pair of spherical mirrors provides a telephoto system for reimaging the focal plane at F/200 onto the lenslet array. The lenslet array samples the focal plane and produces a grid of "spots" or micropupils which are each an image of the telescope pupil. The only aberrations affecting the image quality of the field are from elements in front of the lenslet array. The choice of an all-reflective design for the reimaging system was made to minimize chromatic aberrations that can degrade speckle suppression, which is completely dependent on modeling of chromatic shifts of the speckles. Even the filters were moved into the spectrograph portion to remove their effects on image quality. The plane formed by the grid of pupil images from the lenslet array is the input conjugate plane for a fairly standard prism spectrograph. The spectrograph is an all-refractive design with a collimator and camera system based on Petzval lens systems and is detailed in an earlier SPIE paper by Simon Thibault et al.[15]. The filter and prism lie between the collimator and camera in the spectrograph system. The detector finally sits in the spectrograph focus conjugate to the pupil plane from the lenslet array. For the polarization mode, the spectral prism is removed and a Wollaston prism is inserted into the collimated space to separate the 2 polarization states.

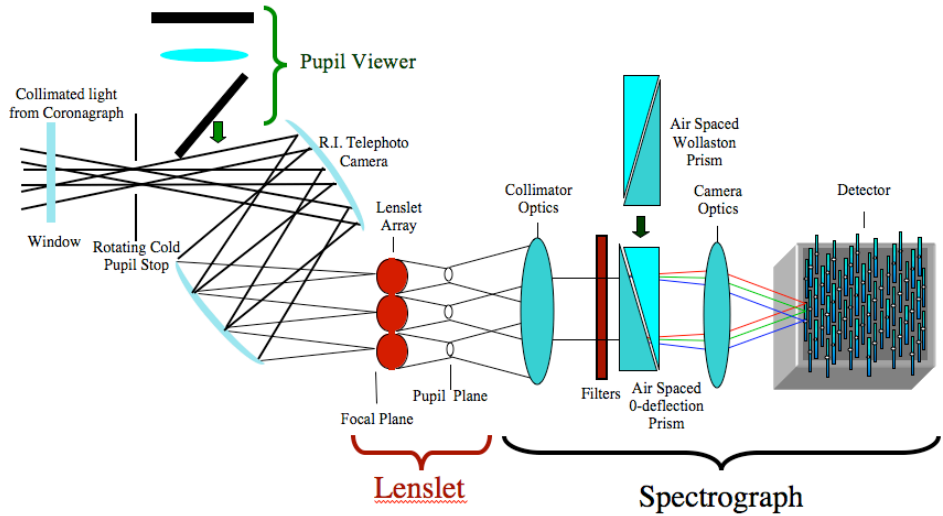

Figure 1. Schematic of Functional Sections of the Optical Design

So three data products can be produced on the detector. With no prisms, the grid of well-separated pupils is directly reimaged onto the detector for an engineering mode. The two science modes shown in Figure 2 use either the Wollaston prism or spectral prism to spread each micropupil. The primary trick is to disperse the pupils so that light from one lenslet does not contaminate light from an adjacent lenslet. The spectral mode is the densest with each spectrum being about 20 pixels long with a fwhm perpendicular to dispersion of about 2 pixels. By separating the dispersed spectra by 22.5 pixels in the dispersion direction and 4.5 pixels between neighbors the spectra remain well separated and can be extracted to compute the full data cube.

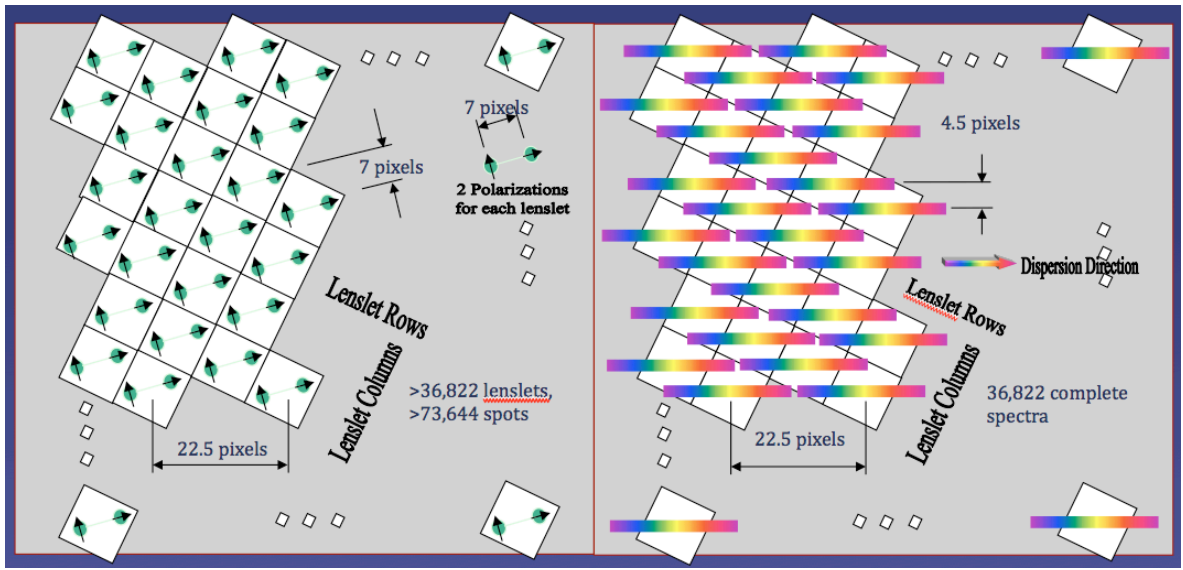

Figure 2. On the left is a cartoon of the array of lenslets (white boxes) mapped onto the detector in the polarization mode. Each lenslet condenses its light into a small dot of light (micropupil) which is then relayed through the spectrograph to the detector. By inserting the Wollaston prism the micropupils are split into their two polarization states with a separation at the detector of 7 pixels. This produces data that is also arranged in a perfect grid but with twice as many dots. By extracting via photometry all of the dots of one polarization states we can produce an image of the field in that state. In a similar fashion, the right panel shows the effect of inserting a dispersive prism. The micropupils are spread into a spectrum less than 20 pixels long in an orientation where micropupils are separated by 22.5 pixels from each other. So each spectrum remains separated on the detector and can be extracted to produce the spectrum of that location in the field. A total of 36,822 lenslets have their complete spectrum fall on the detector giving a field of view of approximately 192x192 field locations.

## 2.1 Lyot stops

In the pupil plane formed by the external calibration unit, there is a wheel with 10 possible Lyot stops. Their design, fabrication and testing are detailed in a previous SPIE paper by A. Sivaramakrishnan et al.[7] An apodized pupil stop and focal plane stop are included in the AO system and the primary purpose of the IFS Lyot stops is to block ringing from the focal plane stop and as a cold stop to block thermal background. The stops also mask off bad actuators in the MEMs mirror that would otherwise scatter light throughout the field. A typical mask is shown in Figure 3. They are aluminum and were aluminum carbon electrode drilled. They are then painted black using Epner Laser Black coating. The different masks use different amounts of oversizing of the spiders, secondary and bad actuators. This is in part to cover the large wavelength range of the instrument, but also for experimentation in contrast under varying conditions. The pupil formed by the CAL unit is 10mm in diameter and the typical clear aperture of the masks is 9.571mm. Even painted black, a small fraction of the blocked light is reflected back down the optical path and could encounter the entrance window, where another small fraction would be reflected back into the spectrograph. To eliminate this source of stray light, the housing of the Lyot mechanism was rotated 1 degree to cause any reflections to miss the window and leave the optical beam. This tilt is shown in the right panel of Figure 3.

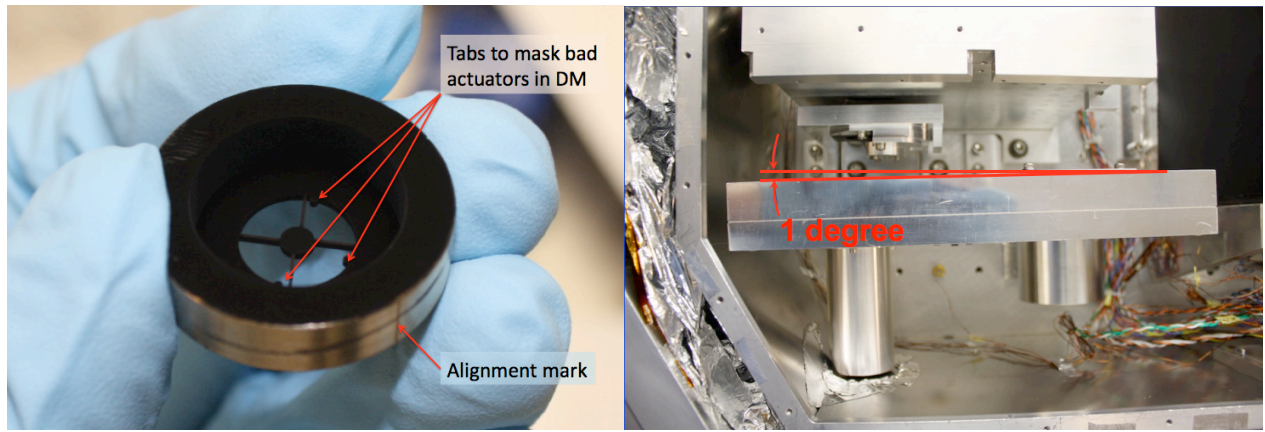

Figure 3. The left panel shows one of the Lyot Masks. In addition to masking off the secondary, outer edge of the primary and the secondary support spiders, three tabs are used to mask bad actuators in the MEMs mirror. An external alignment mark was used to coalign the 9 transmissive masks. To avoid faint ghosts and rejected light from bouncing off the back of the Lyot stops and the dewar window, the entire pupil mechanism is tilted by 1 degree compared to the input beam axis.

## 2.2 Reimaging system

After the Lyot stop, a reimaging system is used to focus the input collimated beam into an image on the lenslet array. With an input pupil diameter of 1 cm and the desire for a 0.014" sampling per 110 micron diameter lenslet, the optics require an F/200 focal ratio and 2 meter focal length. To reduce chromatic aberration as much as possible, a reflective design was selected. The window is the only transmissive optics in front of the lenslet array within the spectrograph. With this very slow focal ratio, the dominant aberrations come from polishing and increase with the number of surfaces. A highly satisfactory solution was found with only two super polished spherical mirrors acting as a telephoto system to package the focal length in the small dewar volume. To further minimize wavefront error due to differential contraction within the glass at cryogenic temperatures, we selected an ultra-low expansion glass for the substrates. Combined with the window, the total estimated wavefront error before the focal plane is imaged at the lenslet array is less than 25 nm rms. This low wavefront error is challenging to measure, but Figure 9 shows the excellent point spread function produced in the final data products from the GPI.

## 2.3 Lenslet array

The lenslet array is the heart of the optical system and serves as the location where the field of view is sampled. Each lenslet concentrates the light from its patch of the sky into a tiny pupil image. These concentrated images are well separated from each other and their spectra can be interleaved on the detector. The lenslet was produced by MEMs optical, which is now part of Jenoptik Optical Systems Inc. They are made with a grayscale lithographic technique that produces a monolithic optic with no internal surfaces. One of the goals of the lenslet design was to minimize the amount of light lost at the gaps between adjacent lenslets. For custom products, MEMS optical can keep these gaps to 2 microns in size. To further reduce their impact, our lenslet design has a very large radius of curvature on the front surface and most of the lens power on the rear surface. The majority of the light from the front surface gap will fall near the corresponding gap on the back surface. So a metalized mask was applied to the back surface over the gaps to further reduce scattered light. Having most of the power in the back surface has an additional benefit that the focus of a lenslet element occurs behind the substrate and the substrate can be significantly thicker than the effective focal length of 578 microns. We selected a 1 mm substrate fabricated out of Infrasil 302. Figure 4 shows a raytrace of three adjacent lens elements within the array. The fill factor of the lenslet array is greater than 95% and the throughput of the Infrasil material with AR coating is greater than 99% per surface.

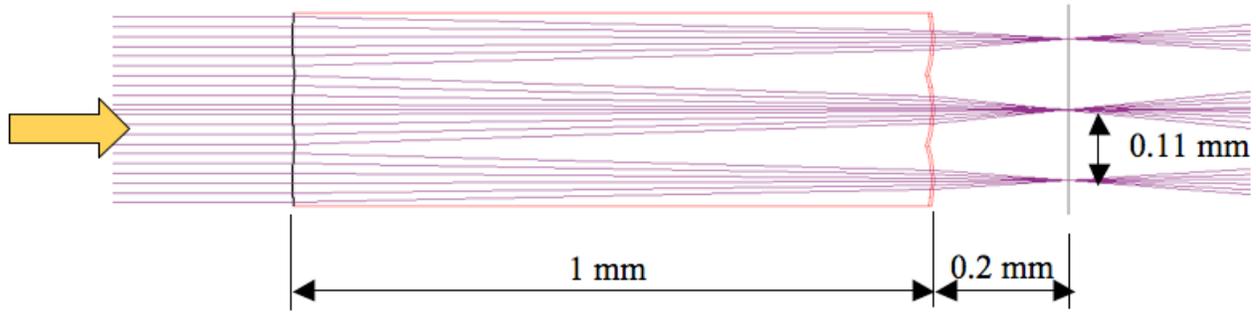

Figure 4 - Ray trace of 3 adjacent lenslets. The light enters from the left side of the figure and the image plane forms on the left suface of the lenslet array. The lenslets' job is to concentrate the light down into well-separated pupil images just after the array. Notice the large change in focal ratio occurring within the array. F/200 light impinges on the lenslet to form the image, and the outgoing beam is F/3.52 including the corners of the square lens elements.

Given the diffraction limited sampling of the lenslets, the size of the pupils is dominated by diffraction effects and geometric aberrations are negligible for the very slow input beam. An early task within the project was to model the effects of this "pupil diffraction". To first order, the beam into the lenslet array is an almost collimated beam into a square optic. So the diffraction pattern at the micropupil plane is dominated by the Fourier transform of a square which is a sinc-squared function in 2-dimensions. The largest pupil size is set by the longest wavelengths where the diffraction is the largest. The spectrograph is then designed to map this largest pupil geometrically to one pixel at the detector. Aberration of the spectrograph optics blurs this into a roughly 2-pixel fwhm spectral feature and sets the spectral resolution and cross talk between the spectra.

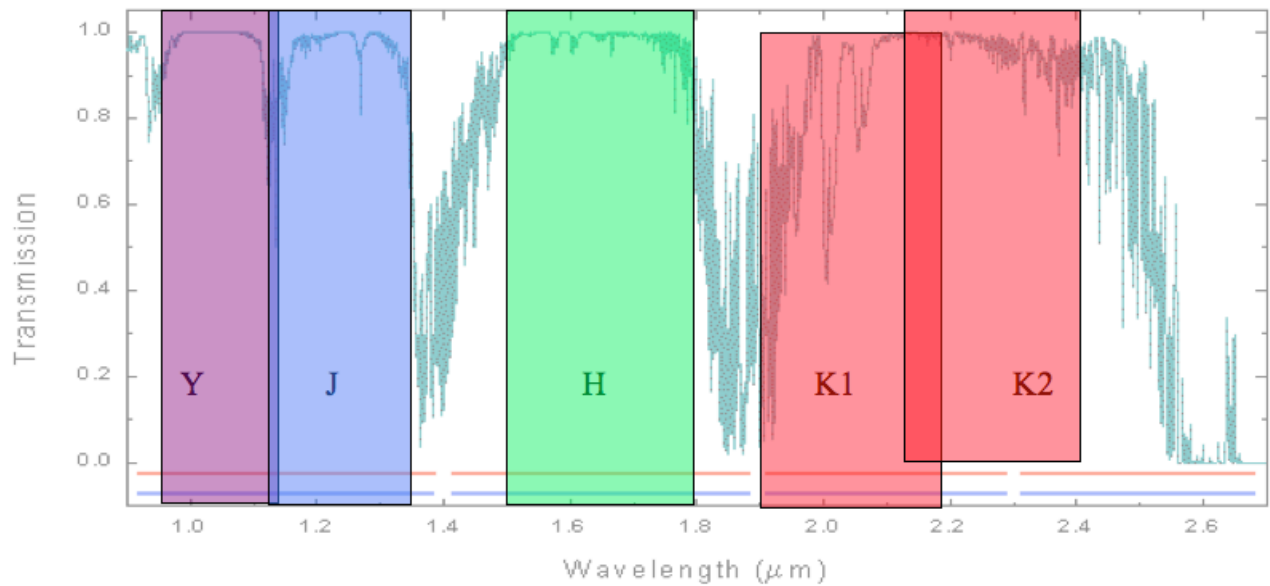

Figure 5 - Atmosphere transmission with approximately 1.6 mm of precipitable water vapour overlayed with the GPI spectrograph filter complement. Due to the higher prism dispersion at long wavelengths, the K-band is split into 2 filters.

### 2.4 Filters

The spectrograph is a low dispersion instrument working from the ground. The filter selection was dominated by the available atmospheric windows and the desire to cover as many wavelengths as possible. Figure 5 shows the ATRAN (Lord, S.D. 1992)[17] model for the atmospheric transmission on Mauna Kea at an airmass of 1.0 and a water vapor column of 1.6 mm. With a prism design optimized in the H-band at a resolution of 45, we found that the dispersion increased at longer wavelengths and forced us to split the K-band into a short and long pair of filters we refer to as K1 and K2. This splitting of the band is partially compensated by having spectral resolutions of approximately 70 and 80 within these filters. The bandpasses, expected spectral resolutions and number of spectral channels for each filter are given in Table 1.

### 2.5 Spectrograph optics

The optics between the lenslet array and the detector form a standard spectrograph with a collimating set of lenses, a dispersive prism and a camera set of lenses. Because the lenslet arrays are quite fast and this is a fully cryogenic system, the optics were relatively challenging. University Laval (ULaval/ImmerVision) and University of Montreal (UdeM) designed the spectrograph as a stand-alone set of optics mounted on its own subplate. They were assembled, aligned and cryogenically tested using an engineering grade H2RG at the University of Montreal prior to delivery to UCLA. The design of the spectrograph is given briefly below and was detailed in a previous SPIE paper[15].

The spectrograph collimator has a focal ratio of F/3.52 (to the corner of the square beam from the lenslet elements) with a 22.44 mm square field. The spectrograph camera has a focal ratio of F/5.89 and remaps the lenslet array onto the science detector. Both are all-refractive except for 2 fold mirrors used for packaging. The 8 lenses include 3 of Cleartran, 4 of BaF2 and 1 of S-FTM16. The estimated throughput of the system is greater than 70% at all wavelengths. The worst component is the single lens of S-FTM16, which has a 93% throughput at 2.4 microns for its thickness.

### 2.6 Spectral Prism

To simplify alignment of the system and to support two dispersive modes and an engineering undispersed mode, both the spectral prism and the Wollaston prism were designed to have no net deflection at a wavelength of 1.65 microns. So the spectral prism uses a pair of prism elements to produce the required R=45 primary spectrum but with no net deflection at this central wavelength. After a significant trade study we settled on the spectral prism pair consisting of BaF2/S-FTM16. The prism is rotated by 6.3 degrees compared to the optical bench in order to disperse along rows of the detector.

### 2.7 Polarization mode

Polarization observations are possible by removing the primary prism and inserting a Wollaston prism. A waveplate is located within the calibration unit and works in conjunction with the prism. The spectrograph optics magnify the lenslet array onto the detector so one lenslet element is 10 pixels across, or just over 14 pixels along the diagonal. Maximum separation of the polarization states is achieved by dispersing them by only 7 pixels at an angle of 45 degrees compared to the lenslet pattern. We selected MgF2 for the Wollaston material for its low cost and widespread availability. Due to the requirement to operate at cryogenic temperatures, an air-spaced Wollaston prism was used to avoid thermal stresses between the two prism elements. The full field is maintained in polarimetry mode, but no significant spectral dispersion remains.

## 3. MECHANICAL DESIGN

The GPI IFS mechanical design relies heavily on heritage developed from previous instruments. It contains five cryogenic mechanisms based on previous generations of mechanisms from the UCLA Infrared Laboratory. It is entirely cooled with a pair of Stirling cycle mechanical refrigerators (CCR) from Sunpower Inc.. The dewar is a six-sided welded aluminum box with one-inch thick walls and a similar cover plate. Precision reference surfaces are welded inside and out to connect to bipods from the rest of GPI, and to the inner optical plate, respectively. After welding, the entire dewar was placed inside a 4-axis computer controlled milling machine and the reference surfaces were cut in a single coordinate system with 25 micron accuracy. The optical bench is a two-inch thick thermally stabilized aluminum plate. The thickness of the plate was set to minimize all differential motion of optical elements under a changing gravity vector. A thin sheet metal plate was bent into an interior 6-sided box and welded to the edge of the optical plate to form the primary cold structure. A thin plate forms the cover of the internal box and forms an almost complete cold shield for all optical elements. Refractive elements after the lenslet array were designed and fabricated at Immervision and the University of Montreal and mounted to a subplate that kinematically mounts to the primary optical bench along with all of the mechanisms, other optics and the detector. The cold structure was mounted to the three reference surfaces on the dewar walls with 12 mm thick G-10 fiberglass struts for thermal isolation. The arrangement of internal and external mounting pads allows differential contraction of the bench to be centered on the lenslet array location. The cold structure of the bench and sheet metal walls along with the inner dewar walls were wrapped with several layers of aluminized mylar for radiation isolation. At operating temperatures, the thermal load is less than 20 Watts.

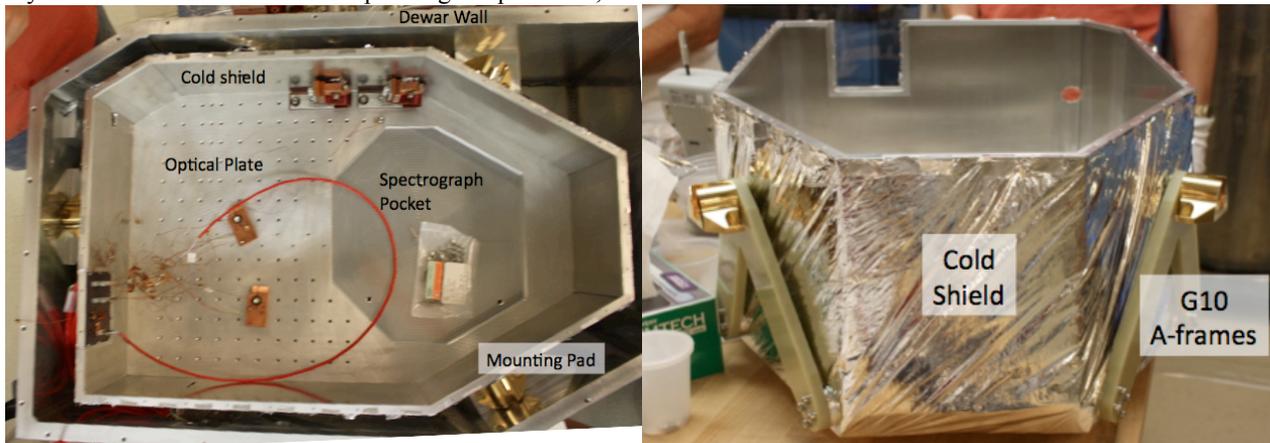

Figure 6. Left panel shows the unpopulated optical bench with pocket hollowed out for the subplate that support the refractive spectrograph elements. The cold shield and its mounting brackets to the dewar wall are shown in both panels.

### 3.1 Closed-cycle cooler

The entire IFS optical bench is operated at a temperature close to 77K. Gifford McMahan (GM) closed cycle refrigerators (CCR's) are the most common cooling method of other Gemini instruments and the one UCLA has used in the past. But these require large shifts in internal helium pressure (typically 50 to 300 PSI) that create a broad spectrum of vibration. Alternative coolers were investigated and initially the team selected a pulse tube cooler from Chart Industries/Q-Drive. Preliminary values from the company suggested that the vibration was more than an order of magnitude below those from a comparable GM cooler and that the majority of the remaining vibration was in two balanced motors that could be separated from the dewar. But the company was unable to deliver a cooler at that time and after waiting approximately a year, we were forced to go with our second choice of a Stirling cycle cooler from Sunpower Inc. The Stirling coolers had an additional benefit of maintaining constant cooling power in any orientation making them ideal for the Cassegrain focus position at Gemini.

While the Sunpower CCRs offer excellent reliability and cooling power, and their total vibration is low compared to a GM cooler, the vibration that is produced is at 60 Hz and produces resonant behaviors within the instrument and the entire observatory at 60 Hz and higher harmonics. This has proven challenging to damp and control. The CCRs are each mounted on a support tower isolated by thick rubber stacks of washers and a flexible bellows to the dewar. Passive dampers are also mounted on each CCR and in strategic places around the instrument. But vibration remains a major concern for GPI and trefoil aberration is seen from the primary mirror that now oscillates due to the CCRs.

The vibration was identified as a problem before shipping from UCLA and a series of mitigation strategies were taken including the isolation and damping mentioned above. But there was still concern about vibration and we had one catastrophic failure of a CCR snout that allowed a release of Helium gas into the dewar while at cryogenic temperatures. So a backup strategy was implemented by cutting an additional port in the dewar wall to potentially mount a pulse tube cooler or other future CCR system. This port was cut and sealed with great care while all optics and electronics were already inside the dewar. The spare port, dual CCRs and their support hardware are shown in Figure 7.

At the time of this writing, Lisa Poyneer has been able to use the AO control system to filter out 60 and 120 Hz components of the measured wavefront and has reduced the impact of the CCR vibration to GPI[18]. But there remains concern about mechanical problems that might arise from the vibration and image problems induced in the other instruments.

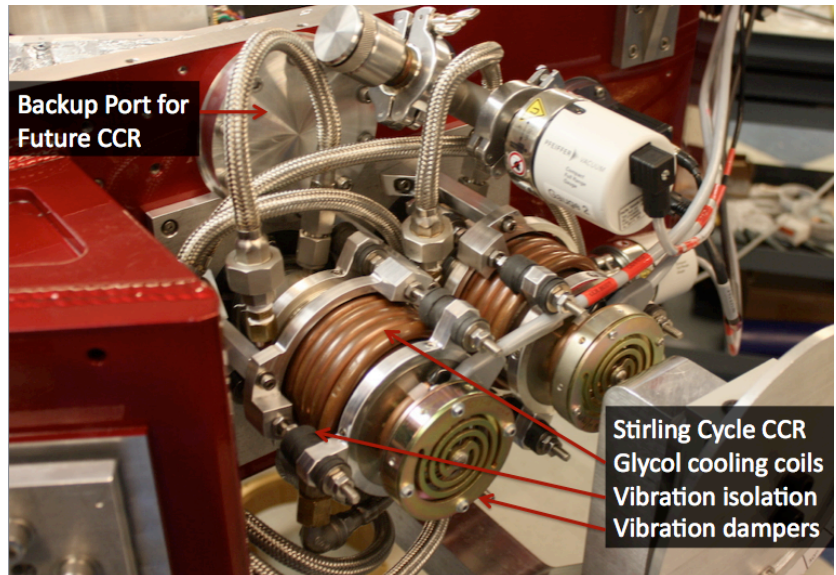

Figure 7. Side of the spectrograph dewar showing dual Stirling cycle refrigerators and extra port for potential future CCR. The Stirling refrigerators are cooled by flowing glycol passed a heat rejection snout and around their bodies with a custom copper coil. Passive dampers reduce their 60Hz vibration and they are supported in a support tower with rubber isolation washers and a metal flex bellows to the dewar.

### 3.2 Mechanisms

The spectrograph has five internal cryogenic mechanisms including a focus drive for the detector that is normally disabled. The four active mechanisms are based on either a wheel or simple slide mechanism and have an extended heritage at UCLA in instruments such as OSIRIS, MOSFIRE, NIRSPEC and GEMINI. All mechanisms use off-the-shelf stepper motors prepared for cryogenic operations at the UCLA lab. Similar motors have demonstrated lives over a decade within existing instruments built at UCLA. Motion control is provided by a Galil DMC-2183 controller. These are identical to the Galil controller cards used throughout GPI.

Lifetime Testing – A failure of any of the 4 primary mechanisms would result in a lengthy shutdown of the instrument (complete warm-up, service and cool-down). So we created a spare wheel and linear stage identical to the core elements in each of the mechanisms and performed lifetime testing. After a simulated 10 years of life they were warmed and inspected for wear. Despite these early tests, minor modifications were made during the integration and testing phases, and a problem with incorrect compression of drive train elements in the two slides was not fully resolved until January 2014 after the first light observations.

Lyot Wheel – A basic wheel in the pupil plane with 10 positions containing 9 Lyot masks and a blank position. Each mask sits in a substrate with an outer diameter of 25 mm and a possible thickness of 6mm. Microswitches are used to uniquely encode each position and verify proper wheel movement. A roller detent is used to hold each wheel mechanism in the final positions.

Pupil Viewing Stage – A two position stage to insert and remove a pick-off mirror to direct light into the pupil viewing camera. This stage is after the Lyot wheel and before the lenslet array. It is out of the beam for normal operation. A worm drive is used to prevent the mechanism from moving under gravity when motor power is removed. A magnetic catch further helps to hold the stage against reference surfaces at each end of travel.

Filter Wheel – A wheel mechanism containing 5 filter positions is located between the spectrograph collimator and camera. The final size of the filters is 50 mm in diameter and the wheel can handle 60 mm filters.

Prism Stage – A two-position stage between the spectrograph collimator and camera to insert either the spectral prism or the Wollaston prism into the beam. Like the pupil viewing stage, a worm drive is used to prevent back-driving and magnets hold the stage against hard stops for repeatability.

## 4. ELECTRONICS DESIGN

The majority of the GPI electronics consist of heritage commercial off the shelf (COTS) hardware and the Teledyne detector suite. All of the GPI electronics are mounted within two nineteen inch racks mounted at the sides of the instrument on the telescope and the IFS electronics are grouped together at the top of one rack.

### 4.1 IFS External Interfaces

The IFS electrical interfaces to the remainder of the GPI instrument consist of Ethernet connections, RS-232 connections and AC power connections. There are two separate local area networks that interface to the IFS, the Control LAN (Local Area Network) and the Data LAN. The LAN's are Ethernet IEEE 802.3 running at one gigabit per second. There are two RS-232 interfaces to the Gemini terminal server, one from the IFS target computer and the other from the motion control unit which is controlled by the overall GPI control computer. There are two separate electrical power interfaces from the IFS electronics to the GPI power distribution system. The two interfaces are both one hundred and twenty volts nominal RMS alternating current power at a frequency of fifty Hertz. The IFS target computer, terminal server, vacuum and temperature controllers are all connected to the uninterruptible power service while the motion control unit utilizes the normal service.

### 4.2 Spectrograph Detector

The spectrograph uses an HAWAII-2 RG array from Teledyne Imaging Sensors. It has a 2048x2048 pixel format with HgCdTe semiconductor material sensitive to wavelengths from 0.8 to 2.5 microns and a quantum efficiency above 75% in the J, H and K-bands. The detector has 32 discrete outputs and with our clocking patterns is completely read out in 1.455 seconds. The IFS detector has a measured dark current below 0.05 electrons per second (this could include all sources of signal within the spectrograph including potentially light leakage). A simple exposure (correlated double sampled) is formed by subtracting two frames and produces an image with a readnoise of 17 electrons. Science operations are achieved by resetting the array and then continuously reading it out for the duration of the exposure. This is termed up-the-ramp sampling and an algorithm combines the frames to produce a single image with multiplicative weights for each frame depending on its position in the sequence and optimized for the readnoise and dark current performance. For 64 frames (about 95 seconds of clock time) the noise (from all sources) in the final image is below 5 electrons per pixel. Since each spectral location of a final data cube is produced from 2-3 pixels in the detector, the final noise per channel in reduced cubes is closer to 7-8 electrons.

### 4.3 Pupil Viewing Camera

Pupil viewing is achieved by inserting a fold mirror and simple reimaging lens after the Lyot wheel mechanism (see Figure 1). The beam is passed out through a window in the side of the dewar and into a commercial InGaAs camera. The model is SU320kTx-1.7RT from Goodrich Sensors Unlimited. The camera has 320x240 pixels with a 40-micron pitch. It has a cut-off wavelength of 1.7 microns, and we use an H-band filter to limit its bandpass to 1.5-1.7 microns. The camera utilizes an industry standard Camera Link interface to a model PCIe DV C-Link frame grabber from Engineering Design Team Inc. (EDT). The frame grabber supports real time image display with an up to 800 megabytes per second DMA channel to the IFS host computers' PCI Express bus.

## 5. SOFTWARE

The spectrograph software system is a subset of the Gemini Planet Imager Automated Software system[19] and is server-client based and divided into hardware-specific subcomponents. For each piece of hardware, there is a subserver dedicated to interfacing with it. All subservers are based on SUN/ONC Remote Procedure Calls (RPC) and written in C, except for the detector server, which mixes in some C++ and C# to interface with the SIDECAR software provided by

Teledyne. All subservers have the same basic structure with the detector server having some extra complexity discussed below. The overall control system for GPI is operated by the top level computer (TLC) and interfaces to the spectrograph through a master server program that also serves as a client to each hardware subserver. Figure 8 shows the overall IFS software system architecture.

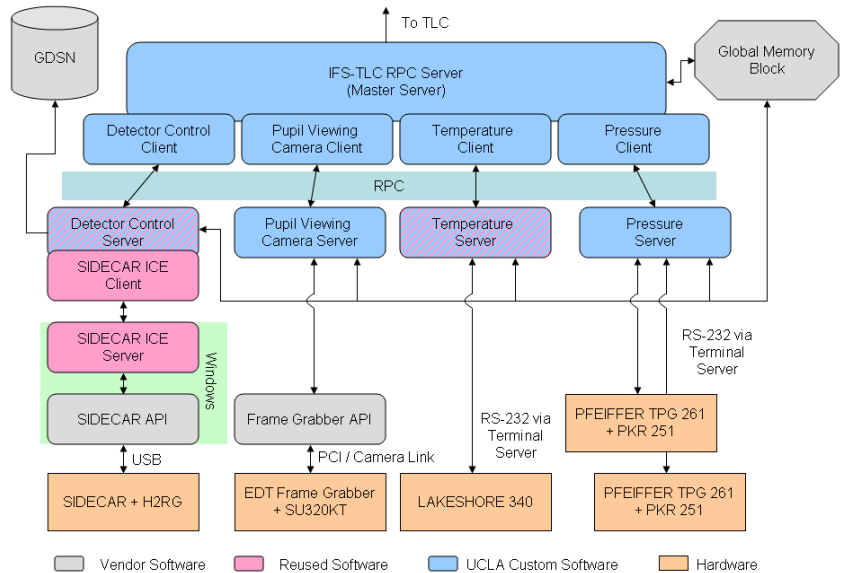

Figure 8- IFS Software Architecture

There are two exceptions to this intermediary role for the master server. First, all subservers have direct access to the Global Memory Block (GMB) for the reading and writing of status and other information made available to other GPI systems. And second the detector server and pupil viewing servers each have a direct link to the Gemini Data Storage Network to write their FITS data.

The science detector software is responsible for interfacing with the Teledyne SIDECAR ASIC electronics that control the infrared focal plane array. The science detector is a Teledyne Scientific Hawaii-2RG. This 2K x 2K HgCdTe detector is supplied with an application specific integrated circuit (ASIC) for control and read out. The ASIC uses a proprietary interface to a module called the "Jade2", also supplied by Teledyne. The Jade2 provides power conditioning and a USB 2.0 compliant interface to the detector target computer. All of the detector control communications and readout data are transferred over the USB 2.0 interface. Teledyne uses a third party USB driver (Bitwise Systems QuickUSB) and their own hardware abstraction layer (HAL) which translates application commands into driver specific commands, hiding USB-specific driver detail from a higher-level application layer. At the time of GPI development the HAL was only compatible with Microsoft .NET web services so we developed an interface with the Internet Communication Engine (ICE) between a Windows machine and the rest of the Linux computers. ICE is designed to be a lightweight alternative to CORBA, providing inter-connectivity among many languages, including C++, C#, Java, and Python, across multiple operating systems, including Windows, Solaris, and Linux. ICE is open-source, free, and released under the terms of the GNU General Public License (GPL). A dedicated windows computer termed the "brick" runs the ICE server that communicates to the LINUX subsystem server for the detector.

A dedicated data reduction pipeline and quicklook viewing program was also developed and often expanded for the spectrograph. Its been presented in a series of papers starting with a description[16] and first test results[10] from Maire et al, in 2010 and won't be discussed further here.

## 6. PERFORMANCE

Laboratory performance of the IFS was previously presented by Jeffrey Chilcote[8], and first light of GPI as a whole was presented by Bruce Macintosh[20]. A variety of results are also presented at the 2014 SPIE conference including in the pipeline and calibration papers. Here we only present a few images demonstrating the final performance on sky. The first image shows a stellar image without the coronagraphic stop illustrating the high quality of the point spread function of GPI as a whole. The estimated H-band Strehl ratio of early images was 89% and a 5-sigma contrast of $10^{-6}$ was achieved at a separation of 0.75 arcseconds.

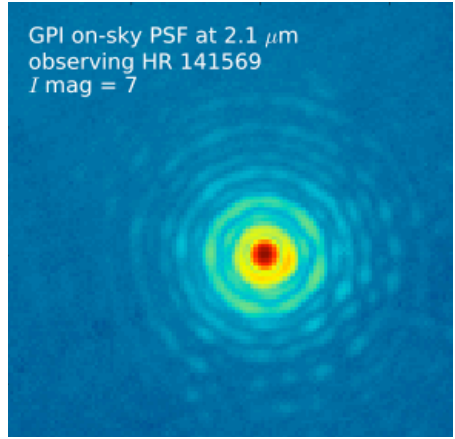

Figure 9. The image above shows a single slice from a datacube taken of a 7th magnitude star at a wavelength of 2.1 microns. Portions of at least 11 Airy rings can be identified illustrating not only the excellent correction of the entire GPI system, but also the low wavefront error of the spectrograph.

In Figure 10 below we present the raw and reduced data from 15 single one-minute exposures on the young nearby star Beta-Pictoris. The left panel shows a few thousand of the raw spectra from a small portion of the detector in one exposure. You can discern the clean separation of the individual spectra. The panel on the right is the reduced image of the field around the star showing the companion planet Beta-Pictoris b located 0.434±0.006 arcseconds from the host star. A complete H-Band spectrum is also present in these data and is being published separately by Jeffrey Chilcote[21].

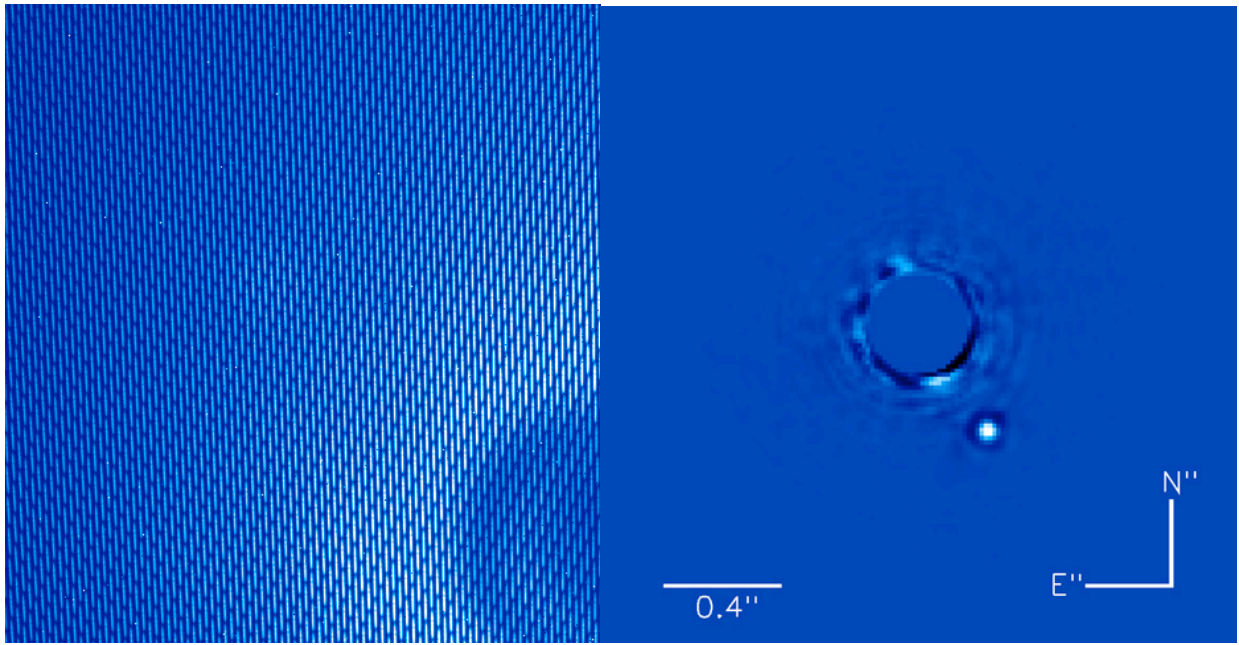

Figure 10. The left panel shows a zoom-in of individual spectra within the halo of Beta-Pictoris from a one-minute exposure. The dark circular region to the bottom left is due to the coronagraph blocking the very center of the star. The right panel is the result of 15 separate one-minute exposures taken on Beta-Pictoris during the first light observing run in November 2013. Differential motion of the sky during the observations, and the color information within the data cube was used to further suppress the stellar halo to reveal the planet Beta-Pictoris b 0.434±0.006 arcseconds to the Southwest of the star. This corresponds to a physical separation of 9 astronomical units.

## 7. SUMMARY

We have presented the design and implementation of the integral field spectrograph for the Gemini planet imager. Several aspects of the design were tailored specifically for its role as a planet finder. These include using a lenslet array style integral field spectrograph to sample the focal plane very early and minimize the impact of internal optics on image quality. The internal optics before the lenslet array are all reflective and include only two highly polished surfaces. Filters were moved behind the lenslet array, and the lenslets were optimized to minimize uncontrolled light from gaps and to put the micropupils outside the lenslet substrate. The Lyot stops has custom tabs to block light from broken actuators in the deformable mirror from scattering light throughout the field and the stops are tilted compared to the beam to direct any weak reflections out of the optical system. The shapes and sizes of the lenslet micropupils are dominated by diffraction and this forces a greater level of magnification onto the detector for spectral interleaving. Dispersions and orientations of the spectral and Wollaston prisms were optimized to separated the spectra on the Hawaii-2RG detector. Spectrograph rms wavefront error is likely below 25 nm but is too low to be easily measurable.

## ACKNOWLEDGEMENT

We would like to thank the staff of the Gemini Observatory for their collaboration on the design of our instrument. The Gemini Observatory is operated by the Association of Universities for Research in Astronomy, Inc., under a cooperative agreement with the NSF on behalf of the Gemini partnership: the National Science Foun- dation (United States), the Particle Physics and Astronomy Research Council (United Kingdom), the National Research Council (Canada), CONICYT (Chile), the Australian Research Council (Australia), CNPq (Brazil), and CONICET (Argentina). Portions of this work performed under the auspices of the U.S. Department of Energy by Lawrence Livermore National Laboratory under Contract DE-AC52-07NA27344.